\documentstyle[aps,eqsecnum,epsf,floats]{revtex}

\begin{document}

\begin{flushright} JLAB-THY-98-46 \\
IU/NTC 98-15 \\
hep-ph/9810418
October 1998 \\ 
\end{flushright} 

\begin{center}
{\Large \bf 
Sudakov Suppression in the Soft Heavy to Light} \\
{\Large \bf  Meson Transition Form Factor.}
\end{center}
\begin{center}
{\sc Adam P. Szczepaniak}
\\
{\em Physics Department and Nuclear Theory Center, \\ 
    Indiana University,
   Bloomington, IN 47405, USA }  \\ 
   
   \vspace{0.2cm}
   
{\sc Anatoly Radyushkin\footnotemark} \\ 
{\em Physics Department, Old Dominion University,}
\\{\em Norfolk, VA 23529, USA}
 \\ {\em and} \\
{\em Jefferson Lab,} \\
 {\em Newport News,VA 23606, USA}
\end{center}
\vspace{2cm}

\footnotetext{Also Laboratory of Theoretical  Physics, 
JINR, Dubna, Russian Federation}

\begin{abstract}

 We calculate hard gluon contribution to the decay vertex that determines 
 the heavy-to-light meson transition form factor at large recoil. 
 It is found that resulting Sudakov suppression significantly decreases the soft, 
wave function  dependent contribution to the form factor. 
 Phenomenological implications of these findings are discussed.

\end{abstract}

\newpage

\section{Introduction}
 Weak decays of heavy mesons give a unique opportunity for studying  
 strong interactions. Due to interference between strong and weak 
 processes,  hadronic matrix elements may be extracted with 
 a much higher precision than from purely strong decays.
 Furthermore, the presence of a large scale provided by the heavy quark mass, 
 $m_Q \gg \Lambda_{QCD}$,  results
 in significant simplifications of the theoretical analysis. 
 For example, the decays in which  another heavy quark with small momentum  
 is produced in the final state can be studied using a nonrelativistic 
 approximation, via an expansion in powers of $\Lambda_{QCD}/m_Q$. 
 On the other hand, highly relativistic decays, with small mass hadrons 
 produced in the final state,  are amenable to perturbative QCD methods 
 as the running coupling constant, $\alpha_s \sim \alpha(m_Q)$ becomes small  
  for highly virtual relativistic quarks.
 In this paper,  we discuss the Sudakov suppression of  the 
 pseudoscalar heavy-to-light meson transition amplitude associated with 
 the production of a relativistic  light quark
 which is near energy shell. 
 If one neglects final state interactions, then  the 
 strong interaction contribution to 
 a two body decay of a heavy meson into two light mesons
(say,  $B \to \pi\pi$), 
 or to a semileptonic decay (such as $B \to \pi l\bar\nu$)   
 is given in terms of the heavy-to-light meson transition matrix element 
  of the weak current.

 For a $J^P=0^- \to 0^-$ transition,  one needs to know
 (both 
for the two-body hadronic
decay and  for  the semileptonic decay at large recoil)  the  magnitude 
of $f_+ = f_+(0)$, the form factor $f_+(Q^2)$ in the limit when the  
  momentum transfer squared   vanishes. 
 The definition of $f_+(Q^2)$  is given by 
  \begin{equation}
 \langle P_H|\bar Q \gamma^\mu q| P_L \rangle = 
f_+(Q^2)(P_H+P_L)^\mu + f_-(Q^2)(P_H-P_L)^\mu.
 \end{equation}
Here $P_H$ and $P_L$ are the momenta of the 
heavy and of the light meson in the final state and $Q = P_H - P_L$ is the 
 four momentum transfer. 
The momenta $P_H$, $Q$ and $P_L$ satisfy, 

\begin{equation}
0 \sim M_L^2  = P_L^2 \  \  , \  \  Q^2 \ll  P_H^2 = M^2_H =
m_Q^2\left(1 +  O \left({{\Lambda_{QCD}}\over m_Q}\right)\right), \label{kin}
\end{equation}

\noindent with $M_L$ and $M_H$ representing the light and the heavy meson 
 masses, respectively and  $m_Q$ being the mass of the decaying heavy quark. 
  The discussion of 
 the form factor in the kinematic region of  
 of Eq.~(\ref{kin})  was given in Ref.~\cite{SHB}. In analogy with  
 exclusive amplitudes at high momentum transfer,   it was shown that 
 exchange of hard gluons  between the decaying heavy quark and the 
 light valence spectator may be required to correlate the produced light 
 quark and  the light spectator in order to enhance the probability for 
 hadronization into a single light meson.  Even though
 perturbative QCD analysis of the heavy-to-light meson transition form factor 
 $f_+(Q^2)$ for $Q^2 \lesssim \Lambda^2_{QCD}$ and an exclusive amplitude at 
 high momentum transfer, $Q^2 \gg  \Lambda^2_{QCD}$ ({ e.g.,} the pion elastic 
 form factor) are  very similar at first  sight, there is a fundamental 
difference  between the two.  In the latter case, 
 the soft  contributions are
$\Lambda^2_{QCD}/Q^2$   suppressed in the asymptotic 
region compared to the  leading 
 perturbative QCD amplitude. On the other hand, 
  in the case of the heavy-to-light meson 
 transitions,  the two contributions are of 
same order in $\Lambda_{QCD}/m_Q$. 
  For this reason, when studying $f_+$, it is crucial to employ 
 a scheme in which both soft and hard contributions are 
  addressed simultaneously\footnote{Sudakov corrections to 
  the hard contribution were considered in Ref.\cite{asy}}. 
 To express $f_+$ 
 entirely in terms of  relativistic hadronic wave functions,  it is necessary 
to quantize the system on the  light front where boosts are kinematical.
 These goals may be achieved using the method of Ref.~\cite{AS,ARJ}. 
 The method is close in spirit to both QCD sum rules~\cite{sum} 
 and to the  
 quark model approach~\cite{qm}. It is based on the analysis of vacuum current 
 correlators. For a given correlator expressed as a sum  
 of perturbative covariant Feynman amplitudes, the integration over the 
 light cone energy, {\it i.e.} the ``minus''  component of the loop momenta 
 is performed analytically. The Borel transformation is then used 
 to model the soft part of the meson-quark vertex (wave function) 
 with the   perturbative gluon exchange kernels explicitly generated 
 in higher-loop diagrams. 
  This is a  novel approach for handling perturbative amplitudes in the 
 off-energy shell environment with nonperturbative hadronic bound states  
 in the asymptotic states. 
 Using this method it was found~\cite{AS} that the gluon exchange 
contribution to $f_+$ may be enhanced  beyond the  collinear 
approximation~\cite{SHB,other}  and that the one loop correction to the 
 decay vertex may  significantly reduce the $O(1)$ soft contribution. 
 In this letter we extend the analysis of Ref.~\cite{AS} 
 to account for the full Sudakov suppression of the decay vertex and calculate 
 the relevant  corrections to $f_+$. 

\newpage

\section{ Soft contribution.} 

Consider the three-point function defined by the correlator
\begin{equation}
\Pi(p_L^2,p_H^2) = (-i)^2\int dx dy e^{ip_Lx-ip_Hy}
  \langle 0| T \bar q(x) \gamma^+ \gamma_5 q(x),
 \bar q(0) \gamma^+ Q(0), \bar Q(y) \gamma^+ \gamma_5 q(y)
| 0 \rangle, \label{tpfplus}
\end{equation}
for $p^+_L = p^+_H \equiv p^+$ where $q$ and $Q$ are the light and heavy 
quark fields, respectively.   
The $O(1)$ contribution to $\Pi$ is shown in Fig.~1a. 
 After a double Borel transformation, we get  

\begin{equation}
\Pi(p_L^2,p_H^2) \to  \Pi(\beta_L,\beta_H) = \left[{1 \over {2\pi i}}\right]^2
\oint dp^2_L dp^2_H \Pi^0(p_L^2,p_H^2) \left[ {{e^{-p^2_L/2\beta_L^2}}  
\over {f_L}} \right]
\left[ {{e^{-(p^2_H-M_H^2)/2m_Q\beta_H}} \over {f_H}} \right], \label{borel}
\end{equation}
where $M_H \sim m_Q + O(\beta_H)$ is the heavy meson mass,  
 $f_{L(H)}$ are the light and heavy meson decay constants and we 
 have neglected the light quark and meson masses, 
 To $O(1)$,  the transformed correlator can be written as 

\begin{equation}
\Pi = 2(P^+)^3 \int  {{d^2{\bf l}_\perp dy}\over {16\pi^3}}
\Psi_H(y,{\bf l}_\perp)\Psi_L(y,{\bf l}_\perp). \label{pi0}
\end{equation}
Here $\Psi_{H(L)}$ may be interpreted as the heavy (light) meson 
light cone wave functions. The Borel transformation is equivalent to the 
Gaussian model and corresponds to

\begin{equation}
\Psi_L = 
{{2\sqrt{6}}\over {f_L}}\exp\left(-{1\over {2\beta_L^2}}
{{{\bf l}_\perp^2}\over {y(1-y)}}\right),\;\;
 \Psi_H = {{2\sqrt{6}}\over {f_H}}
\exp\left({1\over {2m_Q\beta_H}}\left[ M_H^2 
  - {{{\bf l}_\perp^2}\over {y(1-y)}} - {{m_Q^2}\over {1-y}}\right]\right),
\end{equation}
with $-{\bf l}_\perp,(1-y)$ and ${\bf l}_\perp,y$ being the relative light cone 
  momenta of the struck quark and the spectator, respectively.   
 Truncating the phenomenological spectral representation of the correlator 
to a  single contribution from  $J^P=0^-$, $\bar Q q$  heavy and 
$\bar q q$, light meson ground states and comparing with 
Eq.~(\ref{pi0}) leads to the standard light cone representation for 
$f_+=f_+(0)$,

\begin{equation}
f_+ =  \Psi_H \otimes \Psi_L \equiv \int  
{{d^2{\bf l}_\perp dy}\over {16\pi^3}}
\Psi_H(y,{\bf l}_\perp)\Psi_L(y,{\bf l}_\perp). \label{f0}
\end{equation}
If the heavy and light meson interpolating fields in Eq.~(\ref{tpfplus}) 
 are replaced by two identical ones, i.e.,  by either  
  two $\bar Q\gamma^+\gamma_5 Q$  or two $\bar q\gamma^+\gamma_5 q$ currents, 
 then a similar analysis would result in a normalization condition for the meson 
 wave functions, 

\begin{equation}
1 = \Psi_{H(L)} \otimes \Psi_{H(L)}.\label{norm}
\end{equation}
With $f_L=f_\pi \sim f_H = f_B \sim 130\mbox{ MeV}$ this 
 gives  $\beta_H = 300\mbox{ MeV}$, 
and $\beta_L = 400\mbox{ MeV}$, respectively and from Eq.~(\ref{f0}) we obtain

\begin{equation}
f_+ \sim 0.22,
\end{equation}
which is close to the standard value of the soft contribution 
 to the $B\to\pi$ transition form factor, $f_+ \sim 0.3$ obtained 
 using QCD sum rules or more sophisticated quark model wave 
functions~\cite{sum,qm}.

\section{ One loop corrections.}

 The one loop correction to $\Pi$ from the
 dressing of the decay vertex is shown 
in Fig.~1b. In order to ensure current conservation, we also consider 
 one loop corrections to the two propagators connected to this 
vertex.  Integration over the ``minus'' components of loop 
 momenta  picks up poles in the spectator quark and gluon propagators 
 {\it i.e.} puts these two on the mass-shell. 
 The  Borel transformation of Eq.~(\ref{borel}) combined with the 
 the $O(1)$ contribution  from the bare triangle then yields 

\begin{equation}
f_+ = \Psi_H \otimes [I + T] \otimes \Psi_L + \Psi_{gH} \otimes T_{gH} \otimes 
\Psi_L +  \Psi_{H} \otimes T_{gL} \otimes 
\Psi_{gL} +   \Psi_{gH} \otimes I_{gg} \otimes 
\Psi_{gL}  \, . \label{ftot}
\end{equation} 
 The term $T = T(y,{\bf l}_\perp;x,{\bf k}_\perp)$ 
 involving a one gluon exchange  between the valence wave 
functions  is given by 
\begin{eqnarray}
& & T = \phantom{\times}  16\pi^3\delta(y-x)\delta({\bf l}_\perp-{\bf k}_\perp)
 \times \nonumber \\
 & & \phantom{T = } \times \int {{ dz d{\bf p}_\perp} \over {16\pi^3}} 
  {{8\pi\alpha_sC_F}\over {z(1-z)}}\left[
{N(z,{\bf p}_\perp) \over
 {D_H(z,{\bf p}_\perp) D_L(z,{\bf p}_\perp)}}
 -{z\over 2}{1\over D_H(z,{\bf p}_\perp)} - {z\over 2}{1\over D_L(z,{\bf p}_\perp)}
 \right],
 \label{texch}
\end{eqnarray}
with
\begin{equation}
N(z,{\bf p}_\perp) \equiv {{{\bf p}_\perp^2}\over {1-z}} - m_Q^2,\;\;  
D_H(z,{\bf p}_\perp) \equiv {{{\bf p}_\perp^2}\over {z(1-z)}} +{{zm_Q^2}\over {1-z}},\;\;
D_L(z,{\bf p}_\perp) \equiv {{{{\bf p}_\perp^2}\over {z(1-z)}}}. \label{deno}
\end{equation}
 The last two terms in the square bracket in Eq.~(\ref{texch}) come from 
 the loops involving  quark propagators. 
The remaining, three contributions to $f_+$ in Eq.~(\ref{ftot}) involve 
 nonvalence wave functions  $\Psi_g = \Psi_g(y,{\bf l}_\perp;x,{\bf k}_\perp)$,
 which contain  an extra gluon in addition to the two valence quarks. In our approach  
these are given by,   

\begin{eqnarray}
\Psi_{gL} & = & {{\sqrt{8\pi\alpha_sC_F}}\over {D_L(x,{\bf k}_\perp)}}
{{2\sqrt{6}}\over {f_L}}\exp\left(-{1\over {2\beta_L^2}}\left[
{{{\bf l}_\perp^2}\over {y(1-y)}} + {{{\bf k}_\perp^2}\over {x(1-x)(1-y)}}\right]
\right), \nonumber \\
 \Psi_{gH} & = & {{\sqrt{8\pi\alpha_sC_F}}\over {D_H(x,{\bf k}_\perp)}}
{{2\sqrt{6}}\over {f_H}}
\exp\left({1\over {2m_Q\beta_H}}\left[ M_H^2 
  - {{{\bf l}_\perp^2}\over {y(1-y)}} - {{{\bf k}_\perp^2}\over {x(1-x)(1-y)}}
 - {{xm_Q^2}\over {(1-x)(1-y)}}
\right]\right), \nonumber \\
\end{eqnarray}
for the light and heavy meson, respectively.
  The arguments in the exponents 
  are given  
 by the invariant masses of the three body ``valence plus gluon''
  configurations. 
 Finally, the corresponding current matrix elements are given 
 by 

\begin{equation}
  T_{gH} = {{\sqrt{8\pi\alpha_sC_F}} \over {x(1-x)}}
\left[ {x\over 2}- 
{N(x,{\bf k}_\perp) \over {D_L(x,{\bf k}_\perp)}} \right],\;\; 
T_{gL} = {{\sqrt{8\pi\alpha_sC_F}}\over {x(1-x)}}
\left[ {x\over 2}
- {N(x,{\bf k}_\perp) \over {D_H(x,{\bf k}_\perp)}} \right],\;\; 
I_{gg} =  {N(x,{\bf k}_\perp)\over {x(1-x)}}. 
 \label{ts}
\end{equation}
The expression for $f_+$  in Eq.~(\ref{ftot}) is both IR and UV finite even 
 though each individual 
 contribution is divergent. In particular, the first term involving  
the valence quark wave function has an IR double logarithmic divergence 
 coming from the region  $z,{\bf k}_\perp \to 0$,  in which  the two quarks 
 in the vertex loop go on mass-shell. 
To identify contributions from the purely hard gluons,  
 we introduce cut-off functions, $\Theta_i(\mu_i)=\Theta(D_i,\mu_i)$ 
($i=H,L$) where $D_i$ is either 
one of the two denominators in Eq.~(\ref{deno}) 
 such that for small $D_i$, 
 $D_i \ll \mu_i^2$, $\Theta_i(\mu_i) = D_i$ while for  
$D_i \gg \mu_i^2$, $\Theta_i(\mu_i) 
\to 1$.  We may then rewrite Eq.~(\ref{ftot}) as 

\begin{equation}
f_+ = \Psi_H \otimes [ I + T^{hard}]  \otimes \Psi_L + 
 \hat\Psi_{gH} \otimes T^{hard}_{gH} \otimes 
\hat \Psi_L +  \hat \Psi_{H} \otimes T^{hard}_{gL} \otimes 
\hat \Psi_{gL} +   \hat \Psi_{gH} \otimes I_{gg} \otimes 
\hat \Psi_{gL}, \label{fhard}
\end{equation} 
where the gluon exchange kernel staying 
in between the valence wave functions 
 becomes $T \to T^{hard}$
\begin{eqnarray}
& & T^{hard} = 16\pi^3\delta(y-x)\delta({\bf l}_\perp-{\bf k}_\perp)
\nonumber \\
& &  \times \int {{ dz d{\bf p}_\perp} \over {16\pi^3}}
  {{8\pi\alpha_sC_F}\over {z(1-z)}}\left[N(z,{\bf p}_\perp)
{{ \Theta_H(\mu_H)\Theta_L(\mu_L)} \over
 {D_H(z,{\bf p}_\perp) D_L(z,{\bf p}_\perp)}}
 -{z\over 2}{{\Theta_H(\mu_H)}\over D_H(z,{\bf p}_\perp)} - 
{z\over 2}{{\Theta_L(\mu_L)}
\over D_L(z,{\bf p}_\perp)}
 \right], \nonumber \\
\end{eqnarray}
with the remaining terms modified accordingly,
\begin{equation}
 T^{hard}_{gH} = 
 {{\sqrt{8\pi\alpha_sC_F}} \over {x(1-x)}}
\left[ {x\over 2}- 
N(x,{\bf k}_\perp){{\Theta_L(\mu_L)} \over {D_L(x,{\bf k}_\perp)}} \right],\;\; 
T^{hard}_{gL} = {{\sqrt{8\pi\alpha_sC_F}}\over {x(1-x)}}
\left[ {x\over 2}
- N(x,{\bf k}_\perp){{\Theta_H(\mu_H)}\over {D_H(x,{\bf k}_\perp)}} \right]. 
\end{equation}
The modified $\mu_i$-dependent nonvalence wave functions, 
$\hat\Psi_{gi}$ are given by  

\begin{equation}
\Psi_{gi} \to \hat\Psi_{gi} =
\Psi_{gi}(y,{\bf l}_\perp;x,{\bf k}_\perp) - 
{{\sqrt{8\pi\alpha_sC_F}\left[1-\Theta_i(\mu_i)\right]}\over {D_i(x,{\bf k}_\perp)}}
\Psi_i(y,{\bf l}_\perp).
 \label{nonv}
\end{equation}
 Choosing $\Theta_i(\mu_i) = D_i/(D_i + \mu_i^2)$ 
 amounts  to replacing  $D_i$'s with $D_i + \mu_i^2$
 in  the current 
matrix elements. 
 Even though each individual term in Eq.~(\ref{fhard}) becomes now 
$\mu_i$-dependent, after   summation  the $\mu_i$ dependence disappears.  
 Furthermore, for 
  $\mu_L^2  \sim 2\beta_L^2$ and $\mu_H^2 \sim 2m_Q\beta_H$, the second term 
 in Eq.~(\ref{nonv}) 
 strongly reduces the  magnitude of 
 the nonvalence amplitudes $\hat\Psi_{gi}$ 
  and,  therefore, 

\begin{equation}
f_+ \sim \Psi_H\otimes [I + T^{hard}(\mu_i)] \otimes \Psi_L. \label{th}
\end{equation}
 With the above choice of the cut-off functions, the  
 contributions to $f_+$ from nonvalence sectors have been effectively 
 absorbed by the valence sector through mass terms  of the order $\mu_i^2$ 
 added into 
  the free propagators. Having isolated the hard 
contribution to $f_+$, we may 
 sum to all orders in $\alpha_s$ the 
leading single and double logarithms. From Eq.~(\ref{th}) it follows that 
 these are given by

\begin{equation}
f_+ \sim  \Psi_H\otimes S(\mu_i) \otimes \Psi_L,
\end{equation}
with

\begin{equation}
S(\mu_i) = 16\pi^3\delta(x-y)\delta({\bf k}_\perp-{\bf l}_\perp)\left[
1 + {{\alpha_s}\over {2\pi}}C_F
\left( {3\over 4}\log{ {m_Q^2}\over {\mu_L^2}} 
  -{1\over 2}\log^2{{ m_Q^2}\over {\mu_L^2}} + {1\over 2}\log^2{{\mu_H^2}\over {\mu^2_L}}\right)\right].
\end{equation}

\noindent Taking $\mu_L^2 = 2\mu^2$, $\mu_H^2 = 2m_Q\mu$,
after exponentiation this results in

\begin{equation}
f_+ = \left( {{\alpha(m_Q^2)}\over {\alpha(\mu^2)}}
\right)^{-{6\over {33-2N_F}}}\left({{m_Q}\over {\mu}}
\right)^{-{6\over {33-2N_F}}\log{{\alpha(\mu^2)}\over {\alpha(m_Q^2)}}}
\Psi_H \otimes \Psi_L.\label{fin}
\end{equation}

\section{ Conclusions.} 
The first factor in Eq.~(\ref{fin}) comes from the evolution of the leading,  
 UV logarithm which is due to a large difference between the light and 
  heavy quark virtualities~\cite{HQET}. The second term is the 
 Sudakov form factor which suppresses the interaction in the kinematic region 
 where the heavy 
 and light quark go on-mass shell. The important feature of 
 Eq.~(\ref{fin}) is that it factorizes the hard and the soft 
 (i.e., wave function dominated)  contributions. 
 Taking the average value for the factorization scale, $\mu = 500\mbox{ MeV}$ and 
  $m_Q = m_b = 4.8\mbox{ GeV}$,  together with $N_F=4$ and
 $\Lambda_{QCD} = 230\mbox{ MeV}$ we obtain that
   the net result due to hard gluon exchange 
 is a suppression by  approximately 

\begin{equation}
S \sim 0.66
\end{equation}
corresponding to a $34\%$ correction to $O(\alpha^0)$ form factor. 
The standard value for the soft form factor, $f_+ \sim 0.33$ 
is in agreement with the $B\to \pi l\bar\nu$ 
branching ratio as measured recently by CLEO~\cite{cleo}. It also predicts 
the $B\to\pi\pi$ branching ratio to be approximately 
two times smaller than the  current upper limit~\cite{bpipi}. 
Reduction of the form factor by a $\sim 30\%$, as calculated here, 
indicates that contribution from the hard gluon exchange between the 
decaying heavy 
 quark and the light spectator may be comparable to the soft form factor 
 reduced by the Sudakov term. 
In Ref.~\cite{AS} the hard gluon exchange contribution was calculated and 
 indeed found to be approximately $5.3(\alpha_s/\pi)$ times the soft form 
factor. For $\alpha_s \sim 0.3$ this would give additional 40-50\%. 
Therefore  combining the 
  hard gluon exchange contribution with the Sudakov-suppressed soft one 
 increases $f_+$ back to approximately $\sim 0.3$ and agreement 
 with the  semileptonic data is recovered. 
 In other, words, Sudakov effects are compensated 
 by the hard gluon exchange and the net $O(\alpha_s)$
 correction is rather small. 
 The QCD corrections to $f_+$ have also been studied using light cone
 sum rules applied to an off diagonal correlator of heavy-light
 currents taken between vacuum and the light meson
 state~\cite{corr1,corr2}. 
In these approaches
 the soft contribution is expressed as a series over collinear
 terms from operator matrix elements of increasing twist. 
 There the total $O(\alpha_s)$ correction to the twist-2 piece 
 is also small: it does not exceed $20\%$.   
  Our results seem to be somewhat 
 larger. One possible reason being that our approach sums up subleading
 twist contributions in both soft and gluon exchange terms. This
 may be relevant since in the light cone sum rules it is found that
 in the $\alpha_s^0$ order higher twist (3 and 4) contributions  are
as large as the leading one.
A more detailed comparison of the two approaches should be 
undertaken.

\newpage

\section{ Acknowledgments} 
This work was supported by the DOE under grants 
 DE-FG02-87ER40365 (A.S) and 
DE-AC05-84ER40150 (A.R.).

\begin{figure}[htb]
\mbox{
\epsfxsize=5in
\epsffile{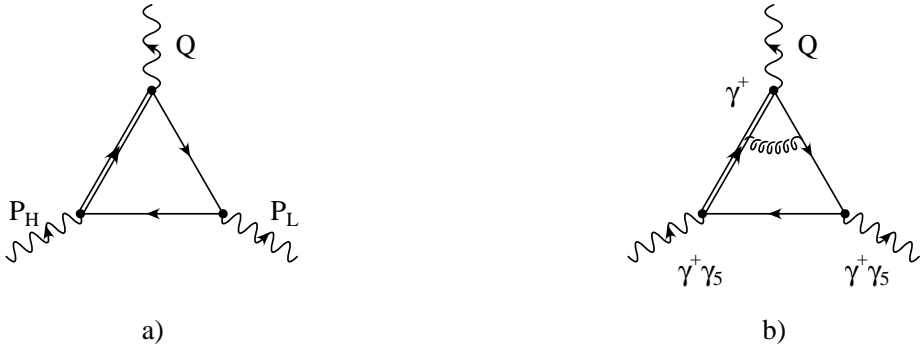}}
\vspace{0.5cm}
{\caption{\label{fig:amb}
  Perturbative expansion of the three point function used to 
 calculate $f_+$.  }}
\end{figure}


\begin{references}

\bibitem{SHB} A.~P.~Szczepaniak, E.~M.~Henley and S.~J.~Brodsky, Phys. Lett. 
{\bf B 243} (1990) 287.

\bibitem{asy}  R.~Akhoury, G.~Sterman and  Y.~P.~Yao,  
 Phys. Rev. {\bf D50} (1994) 358. 
 
 
 \bibitem{AS} A.~P.~Szczepaniak, Phys.Rev. {\bf D54} (1996) 1167. 
  
\bibitem{ARJ} A.~P.~Szczepaniak,  A.~Radyushkin and C.~R.~Ji, 
 Phys. Rev. {\bf D 57} (1998) 2813.

\bibitem{sum}  V.M.~Belyaev, A.~Khodzhamirian and
R.~R\"{u}ckl, Z. Phys. {\bf C 60} (1993) 349; A.~Ali, V.M.~Braun,
 H.~Simma,
  Z. Phys. {\bf C 63} (1994) 437; V.M.~Braun, A.~Khodzhamirian and
R.~R\"{u}ckl,  Phys. Rev. {\bf D 51} (1995) 6177.

\bibitem{qm} M.~Bauer, B.~Stech and M.~Wirbel, Z.~Phys. {\bf C 29} (1985) 637; 
{\bf C 34} (1990) 103;  N.~Isgur, D.~Scora, B.~Grinstein and M.~B.~Wise, 
 Phys. Rev. {\bf D 39} (1989) 799; D.~Scora and N.~Isgur, Phys. Rev. 
 {\bf D 52} (1995) 2783.



\bibitem{other}  G.~Burdman and J.F.~Donoghue, Phys. Lett. {\bf B 270} (1991)
 55; H.~Simma and D.~Wyler, Phys. Lett. {\bf B 272} (1991) 55; 
C.~E.~Carlson and J.~Milana, Phys. Lett. {\bf B 301} (1993) 237; 
R.~Fleischer, Z.~Phys. {\bf C 58} (1993) 483; M.~Dahm, R.~Jakob and P.~Kroll, 
 Z.~Phys. {\bf C 68} (1995) 595.


\bibitem{HQET} M.~B.~Voloshin and M.~A.~Shifman, Sov. J.~Nucl. Phys. {\bf 45} 
 (1987) 292; H.~D.~Politzer and M.~B.~Wise, Phys. Lett. {\bf B 206} (1988) 
681; {\bf B 208} (1988) 504. 



\bibitem{cleo} CLEO Collab., J.~P.~Alexander {\it et al.}, CLNS-96-1419 (1996).

\bibitem{bpipi} CLEO Collab., M.~Battle {\it et al.}, Phys. Rev. Lett. 
{\bf 71} (1993) 3922.

\bibitem{corr1} A. Khodjamirian, R. Ruckl, S. Weinzierl, O. Yakovlev,
    Phys. Lett. B {\bf 410}, 275 (1997). 

\bibitem{corr2} E. Bagan, P. Ball, V.M. Braun, 
    Phys. Lett. B {\bf 417} 154, (1998). 




\end{references}
\end{document}